\pdfoutput=1

\RequirePackage{snapshot}
\documentclass[12pt]{article}

\usepackage[OT1]{fontenc}

\usepackage[utf8]{inputenc}

\usepackage
[a4paper,tmargin=3truecm,bmargin=3truecm,rmargin=2.5truecm,lmargin=2.5truecm,twoside,verbose=true]{geometry}
\usepackage{amsmath,amssymb,latexsym}

\usepackage{stmaryrd,wasysym,upgreek,mathrsfs,dsfont,mathtools}
\usepackage[english]{babel}
\usepackage{graphicx,color}
\usepackage{slashed,subfig}
\usepackage[vflt]{floatflt}

\allowdisplaybreaks[1]

\usepackage[pdftex]{hyperref}
\hypersetup{%
  colorlinks=true,%
  linkcolor=black,%  
  anchorcolor=black,%
  citecolor=black,%
  urlcolor=black,%
  pdftitle={Quantum field theory on the degenerate Moyal space},%
  pdfauthor={Fabien Vignes-Tourneret,Harald Grosse},%
  pdfsubject={Research paper},%
  pdfkeywords={Quantum Field Theory, renormalisation, non-commutative geometry},%
  bookmarksopen=false,%
}

\usepackage[hyperref,thmmarks,amsmath]{ntheorem}
\usepackage{fabmesthm-en}
\usepackage{fabmacromath}

\usepackage{pdfsync}

\newcommand{\lrtimes}{\super{\ltimes}{\rtimes}}

\numberwithin{equation}{section}

\title{\textsf{Quantum field theory on the degenerate Moyal space}}
\author{\textrm{Harald Grosse, Fabien Vignes-Tourneret}}
\date{}

\begin{document}

\maketitle
\vspace*{-1cm}
\begin{center}
\textrm{Working group on mathematical physics\\
  Faculty of physics, university of Vienna\\
  Boltzmanngasse 5,\\
  A-1090 Vienna, Austria\\
  e-mail: \textsf{Harald.Grosse@univie.ac.at, fabien.vignes@gmail.com}}
 \end{center}%

\begin{abstract}
  We prove that the self-interacting scalar field on the four-dimensional \emph{degenerate} Moyal plane is renormalisable to all orders when adding a suitable counterterm to the Lagrangean. Despite the apparent simplicity of the model, it raises several non trivial questions. Our result is a first step towards the definition of renormalisable quantum field theories on a non-commutative Minkowski space.
\end{abstract}

\section{Motivations}
\label{sec:introduction}

For the last five years much has been done in order to determine renormalisable \encv{} quantum field theories \cite{GrWu04-3,Rivasseau2005bh,xphi4-05,RenNCGN05,Grosse2005ig,Grosse2006qv,Grosse2006tc,Langmann2002ai,Langmann2003cg,Langmann2003if,Gurau2008aa}. Nevertheless all the known models are more or less of the type of a \emph{self-interacting} scalar field on a \emph{Euclidean Moyal space}. So it is quite important to extend the list of renormalisable \encv{} models.

The first solution to the uv/ir mixing problem consisted in adding an harmonic potential term to the quadratic part of the Lagrangean \cite{GrWu04-3}. We propose here to test such a method on a $\phi^{4}$-like model on a \emph{degenerate} four-dimensional Moyal space. By degenerate we mean that the skew-symmetric matrix $\Theta$ responsible for the non-commutativity of the space will be degenerate which implies that some of the coordinates will commute.

To understand why we are not addressing a trivial question, let us remind the reader with a precise statement of the problem. We consider scalar quantum field theories on a (degenerate) Moyal space $\R^{D}_{\Theta}$. The algebra of functions on such a space is generated by the coordinates $\{x^{\mu}\}$, $\mu\in\lnat 0,D-1\rnat$ satisfying the following commutation relation
\begin{align}
  \lsb x^{\mu},x^{\nu}\rsb=&\imath\Theta^{\mu\nu}\bbbone
\end{align}
where $\Theta$ is a $D\times D$ skew-symmetric constant matrix. This algebra is realized as the linear space of Schwartz-class functions $\cS(\R^{D})$ equipped with the Moyal-Weyl product: $\forall f,g\in\cS(\R^{D})$,
\begin{align}
  (f\star_{\Theta} g)(x)=&\int_{\R^D} \frac{d^{D}k}{(2\pi)^{D}}d^{D}y\, f(x+{\textstyle\frac 12}\Theta\cdot
  k)g(x+y)e^{\imath k\cdot y}.
\end{align}
In the following we will consider degenerate $\Theta$ matrices which means that $d$ out of the $D$ coordinates will be commutative. In \cite{Wang2008vn} the non-commutative \emph{orientable} $(\bar{\Phi}\star\Phi)^{\star 3}$ model on $\R^{3}_{\Theta}$ has been considered:
\begin{multline}
  S_{6}[\phib,\phi] =\int d^3x \Big( -\frac{1}{2} \partial_\mu \phib
  \star \partial^\mu \phi +\frac{\Omega^2}{2} (\tilde{x}_\mu \phib)\star (\tilde{x}^\mu \phi ) + \frac{1}{2} m^2
  \,\phib \star \phi\\
  + \frac{\lambda_{1}}{2} (\phib\star\phi)^{\star 2}
  + \frac{\lambda_{2}}{3} (\phib\star\phi)^{\star 3}\Big)(x).\label{eq:phi6Omega}
\end{multline}
Being skew-symmetric $\Theta$ is necessarily degenerate in odd dimensions such as three. The main result of \cite{Wang2008vn} is that the complex orientable $(\bar{\Phi}\star\Phi)^{\star 3}$ model is renormalisable to all orders. What about its real counterpart? It is indeed a natural question because the graphs responsible for the uv/ir mixing cannot be generated by such a complex interaction. In an appendix, Z.~Wang and S.~Wan \cite{Wang2008vn} exhibited a first problem concerning the real model. The $\Phi^{\star 6}_{3}$ model on $\R^{3}_{\Theta}$ leads both to orientable and non-orientable graphs \cite{xphi4-05,RenNCGN05}. The upper bound these two authors were able to prove (the power counting) was not sufficient to discard non-orientable graphs. If that bound is optimal, it would remain logarithmically divergent planar two-point graphs with two broken faces. These graphs are non-local and responsible for the now famous uv/ir mixing.

In section \ref{sec:power-counting} we give a strong argument which tends to prove that the power counting given in \cite{Wang2008vn} is actually optimal with respect to the behaviour of the planar two-broken face graphs. We also compute the power counting of our model. The section \ref{sec:definition-model} is devoted to definitions and the statement of our main result. In section \ref{sec:renormalisation} we perform the renormalisation and identify the missing counterterm. Our conclusions take place in section \ref{sec:conclusion}.

\section{Definition of the model}
\label{sec:definition-model}

Motivated by the remarks made in section \ref{sec:introduction}, we want to address the question of the renormalisability of the \emph{real} $\Phi^{\star 4}_{4}$ and $\Phi^{\star 6}_{3}$ models with degenerate $\Theta$ matrix. Whereas we will mainly focus on the $\Phi^{\star 4}$ model, our results apply to the $\Phi^{\star 6}$ case as well.

\subsection{Main result}
\label{sec:main-result}

We are going to prove the following
\begin{thm}\label{thm:MainResult}
  The quantum field theory defined by the action
  \begin{multline}
    S[\phi]=\int_{\R^{4}} d^{2}xd^{2}y\,\frac 12\phi(x,y)(-\Delta+\frac{\Omega^{2}}{\theta^{2}}y^{2}+m^{2})\phi(x,y)\\
    +\frac{\kappa^{2}}{\theta^{2}}\int_{\R^{6}} d^{2}xd^{2}yd^{2}z\,\phi(x,y)\phi(x,z)
    +\frac{\lambda}{4}\int_{\R^{4}}d^{4}x\,\phi^{\star 4}(x)\label{eq:ActionMinim}
  \end{multline}
is renormalisable to all orders of perturbation.
\end{thm}
To this aim, we will treat the new counterterm (the coupling constant of which is $\kappa^{2}$) as a perturbation. Such a counterterm linking two propagators will be called \textbf{$\mathbf{\kappa}$-insertion} or simply insertion. Note that the four-valent vertex in (\ref{eq:ActionMinim}) has the same form as on the non-degenerate Moyal space except that its oscillation only involves the \encv{} directions.

In the following, we use the momentum space representation. Our new counterterm is then given by:
\begin{align}
  \frac{\kappa^{2}}{\theta^{2}}\int_{\R^{6}} d^{2}xd^{2}yd^{2}z\,\phi(x,y)\phi(x,z)=&%
  \frac{\kappa^{2}}{(2\pi\theta)^{2}}\int_{\R^{2}}d^{2}p\,\phih(p,0)\phih(-p,0).\label{eq:FactTermPSpace}
\end{align}
The interaction term reads:
\begin{subequations}
  \begin{align}
    \int_{\R^{4}} d^{4}x\,\phi^{\star 4}(x)=&\frac{1}{\pi^{4}|\det\Theta|}\int\prod_{j=1}^{4}
    \frac{d^{4}p_{j}}{(2\pi)^{4}}\phih(p_{j})\,\,\delta\Big(\sum_{i=1}^{4}p_{i}\Big)e^{-\imath\hat{\varphi}},\label{eq:int-Moyal-Momentum}\\
    \text{with }\hat{\varphi}\defi&\sum_{i<j=1}^{4}p_{i}\wed p_{j}\text{ and }p_{i}\wed p_{j}\defi\frac 12p_{i}\Theta p_{j}\\
    \text{and where }\sum_{i<j=1}^{4}\defi&\sum_{i=1}^{4}\sum_{\substack{j=1\tq\\j>i}}^{4}.\nonumber
  \end{align}
\end{subequations}
From (\ref{eq:int-Moyal-Momentum}) one reads that by convention all the momenta are considered incoming. Note once more that the oscillations only involve the \encv{} directions so that the interaction is local in the commutative directions.

Let $p,q$ (resp.\@ $\kp,\kq$) denote (two-dimensional) momenta in the commutative (resp.\@ \encv{}) directions. Let $\Ot\defi 2\theta^{-1}\Omega$. The propagator corresponding to the quadratic part of \eqref{eq:ActionMinim} is given by:
\begin{align}
  \Ch(p,\kp;q,\kq)=&\frac{\Omega}{\pi\theta}\int_{0}^{\infty}\frac{d\alpha}{\sinh(2\Ot\alpha)}\,\delta(p+q)e^{-\alpha (p^{2}+m^{2})}\,e^{-\frac{\Ot}{4}\coth(\Ot\alpha)(\kp+\kq)^{2}-\frac{\Ot}{4}\tanh(\Ot\alpha)(\kp-\kq)^{2}}.\label{eq:Propag}
\end{align}

\subsection{Feynman graphs}
\label{sec:feynman-graphs}

\begin{figure}[htb]
  \centering
  \includegraphics[scale=1]{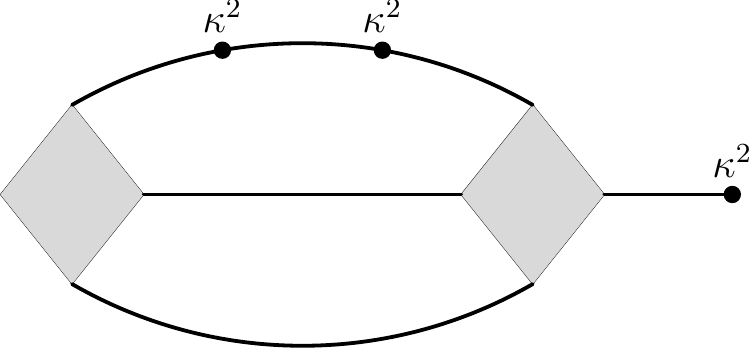}
  \caption{Example of graph with insertions}
  \label{fig:ExGraphIns}
\end{figure}

Let $G$ be a Feynman graph of the model (\ref{eq:ActionMinim}). There are two ways of considering the $\kappa$-insertions in it. Either we think of them as vertices. In this case, $G$ is made of four- and two-valent vertices linked to each other by the \textbf{edges} of $G$. These ones correspond to the propgator $(p^{2}+\Omega^{2}y^{2})^{-1}$. Or we consider that the vertices of $G$ are the only Moyal ones. In this case the insertions belong to \textbf{generalised lines}. These lines are composed of a series of edges related by $\kappa$-insertions. When not explicitly stated, we will always consider a generalised line as a \emph{single} line, whereas it is composed of several edges. In this picture the edges linking two Moyal vertices are called \textbf{simple lines}. Note also that some ``external'' insertions may well appear, see figure \ref{fig:ExGraphIns} for an example.\\

\noindent
We now fix the notations we use throughout this article:
\begin{defn}[Graphical notations]
  Let $G$ be a Feynman graph corresponding to the model \eqref{eq:ActionMinim}. We define:
  \begin{itemize}
  \item $\mathbf{E}(G)$ (resp.\@ $\mathbf{Ex}(G)$) to be the set of \emph{internal} (resp.\@ external) edges of $G$. $E$ is the disjoint union of the sets $\mathbf{E_{0}}$ of internal simple lines and $\mathbf{E_{\kappa}}$ of internal generalised lines. The respective cardinalities of $E, E_{0}$ and $E_{\kappa}$ are denoted by $\mathbf{e},\mathbf{e_{0}}$ and $\mathbf{e_{\kappa}}$.
  \item The number of external lines of $G$ is $\mathbf{N}(G)\fide \card Ex$.
  \item The number of external insertions is $\mathbf{N_{\kappa}}$.
  \item Let $\mathbf{V}(G)$ bet the set of vertices of $G$. We note $v\defi\card V$.
  \item The number of connected components\footnote{Please note that ``connected components'' will also be used to denote the quasi-local subgraphs of $G$ in the framework of the multiscale analysis.} of $G$ is called $\mathbf{k}(G)$.
  \item Given a spanning tree $\mathbf{\cT}(G)$, let $\mathbf{L}(G)\defi E(G)\setminus\cT$ be the set of loop lines in $G$. Its cardinality is $\card L=e-v+k\fide \mathbf{n}(G)$. By analogy, we write $\mathbf{n_{0}}$ (resp.\@ $\mathbf{n_{\kappa}}$) for the cardinality of $L\cap E_{0}$ (resp.\@ $L\cap E_{\kappa}$).
  \end{itemize}
\end{defn}

\subsection{Multiscale analysis}
\label{sec:multiscale-analysis}

We use the multiscale analysis techniques \cite{Riv1}. This means that we first slice the propagator in the following way:
\begin{subequations}
  \begin{align}
    \Ch(p,\kp;q,\kq)\fide&\sum_{i=0}^{\infty}\Ch^{i},\\
    \Ch^{0}(p,\kp;q,\kq)\defi&\frac{\Omega}{\pi\theta}\int_{1}^{\infty}\frac{d\alpha}{\sinh(2\Ot\alpha)}\,\delta(p+q)e^{-\alpha (p^{2}+m^{2})}\,e^{-\frac{\Ot}{4}\coth(\Ot\alpha)(\kp+\kq)^{2}-\frac{\Ot}{4}\tanh(\Ot\alpha)(\kp-\kq)^{2}},\\
    \Ch^{i}(p,q)\defi&\frac{\Omega}{\pi\theta}\int_{M^{-2i}}^{M^{-2(i-1)}}\frac{d\alpha}{\sinh(2\Ot\alpha)}\,\delta(p+q)e^{-\alpha
      (p^{2}+m^{2})}\,e^{-\frac{\Ot}{4}\coth(\Ot\alpha)(\kp+\kq)^{2}-\frac{\Ot}{4}\tanh(\Ot\alpha)(\kp-\kq)^{2}}.
  \end{align}\label{eq:PropSlice}
\end{subequations}
where $M>1$. Each propagator $\Ch^{i}$ bears both uv and ir cut-offs. A graph expressed in terms of these sliced propagators is then convergent. The divergences are recovered as one performs the sum over the so-called \emph{scale indices} (the $i$ in $\Ch^{i}$). Now we only study graphs with sliced propagators. Each line of the graph bears an index indicating the slice of the corresponding propagator. A map from the set of lines of a graph $G$ to the natural numbers is called a \emph{scale attribution} and written $\mathbf{\mu(G)}$.

Certain subgraphs of $G$ are of particular importance. These are the ones for which the smallest index of the internal lines of $G$ is strictly higher than the biggest index of the external lines. These subgraphs are called \emph{connected components} or quasi-local subgraphs: let $G^{i}$ be the subgraph of $G$ composed of lines with indices greater or equal to $i$. $G^{i}$ is generally disconnected. Its connected components (the quasi-local subgraphs) are denoted by $G^{i}_{k}$. By construction they are necessarily disjoint or included into each other. This means that we can represent them by a tree, the nodes of which are connected components and the lines of which represent inclusion relations. This tree is called \emph{\GNt}.

\subsection{Topology and oscillations}
\label{sec:topol-oscill}

Let G be a graph with $v$ vertices and $e$ internal lines. Interactions of
quantum field theories on the Moyal space are only invariant under \emph{cyclic permutation} of the incoming/outcoming fields. This restricted invariance replaces the permutation invariance which was present in the case of local interactions.

A good way to keep track of such a reduced invariance is to draw Feynman graphs as ribbon graphs. Moreover there exists a basis for the Schwartz class functions where the Moyal product becomes an ordinary matrix product \cite{GrWu03-2,Gracia-Bondia1987kw}. This further justifies the ribbon representation.\\

\begin{figure}[htbp]
  \centering 
  \subfloat[$p$-space representation]{{\label{fig:x-rep}}\includegraphics[scale=.8]{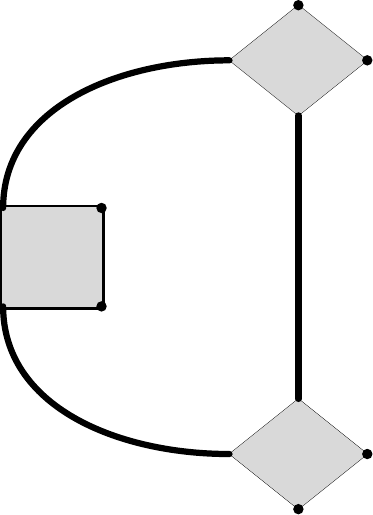}}\hspace{2cm}\qquad
  \subfloat[Ribbon representation]{\label{fig:ribbon}\includegraphics[scale=1]{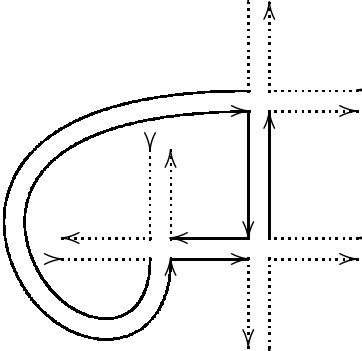}}
  \caption{A graph with two broken faces}
  \label{fig:broken-ex}
\end{figure}

Let us consider the example of figure \ref{fig:broken-ex}.
Propagators in a ribbon graph are made of double lines. Let us call $f$ the number of faces (loops made of single lines) of a ribbon graph. The graph of figure \ref{fig:ribbon} has $v=3, e=3, f=2$. Each ribbon graph can be drawn on a manifold of genus $g$. The genus is computed from the Euler characteristic $\chi=v-e+f=2-2g$ (for an orientable surface). If $g=0$ one has a {\it planar} graph, otherwise one has a {\it non-planar} graph. For example, the graph of figure \ref{fig:ribbon} may be drawn on a manifold of genus $0$. Note that some of the $f$ faces of a graph may be ``broken'' by external legs. In our example, both faces are broken. We denote the number of broken faces by $b$. A graph with only one broken face is called \textbf{regular}.

\subsection{Momentum space representation}
\label{subec:moment-space-repr}

The expression for the oscillation of a general graph that Filk obtained in \cite{Filk1996dm} was based on the assumption that the propagator conserves momentum. When one adds an $x^{2}$ term to the action, the corresponding new propagator breaks translation invariance and so does not conserve momentum. In \cite{RenNCGN05}, one of us computed the expression for the vertex oscillations of a general graph for any propagator. That was done in $x$-space. Here we redo it but in momentum space. Whereas the proof follow the same line we give it for completeness.

\begin{defn}[Line variables]\label{def:LineVar}
  Let $G$ be a graph and fix a rooted spanning tree $\cT$. Let $\ell\in E(G)$ be a line which links a momentum $p_{\ell_{1}}$ to another one $p_{\ell_{2}}$. When turning around the tree $\cT$ counterclockwise, one meets $p_{\ell_{1}}$ first, say. One defines $\mathbf{p_{\ell}}\defi p_{\ell_{1}}-p_{\ell_{2}}$ and $\delta p_{\ell}\defi p_{\ell_{1}}+p_{\ell_{2}}$.
\end{defn}
\begin{defn}[Arches and crossings]\label{def:ArchCross}
  Let $\ell=(p_{\ell_{1}},p_{\ell_{2}})$, $\ell'=(p_{\ell'_{1}},p_{\ell'_{2}})$ and $p_{k}$ an external momentum. One says that $\ell$ \emph{arches over} $p_{k}$ if, when turning around the tree counterclockwise, one meets successively $p_{\ell_{1}}$, $p_{k}$ and $p_{\ell_{2}}$. One writes then $p_{\ell}\mathbf{\supset} p_{k}$.

Considering the two lines $\ell$ and $\ell'$, if one meets successively $p_{\ell_{1}}$, $p_{\ell'_{1}}$, $p_{\ell_{2}}$ and $p_{\ell'_{2}}$, one says that $\ell$ \emph{crosses} $\ell'$ by the left and writes $\ell\mathbf{\ltimes}\ell'$.
\end{defn}
In the following we call rosette factor of a graph $G$ the complete vertex oscillations of $G$ plus a delta function obtained form the conservation of momentun at each vertex.
\begin{lemma}[Tree reduction]
  \label{lem:TreeReduc}
  Let $G$ be a graph with $v(G)=v$. The rosette factor after a complete tree reduction is
  \begin{align}
    &\delta_{G}\Big(\sum_{i=1}^{2v+2}p_{i}+\sum_{l\in\cT}\delta p_{l}\Big)\,\exp(-\imath\varphi),\label{eq:TreeRedDelta}\\
    &\varphi=\sum_{i<j=1}^{2v+2}p_{i}\wed p_{j}+\sum_{i<\cT}p_{i}\wed\delta p_{l}+\sum_{\cT<\cT}\delta p_{l}\wed\delta p_{l'}.\label{eq:TreeReducOscill}
  \end{align}
\end{lemma}
\begin{proof}
  We prove it by induction on the number of vertices. Let us assume that we have contracted $k-1$ tree lines, $k<v$. These lines form a partial tree $\mathbf{\cT_{k}}$. We now want to reduce the tree line between our rosette $V_{k}$ and a usual Moyal vertex $V_{1}$:
  \begin{align}
    V_{k}=&\delta_{G}\Big(\sum_{i=1}^{2k+2}p_{i}+\sum_{l\in\cT_{k}}\delta p_{l}\Big)\,\exp(-\imath\varphi_{k}),\label{eq:VkDelta}\\
    \varphi_{k}=&\sum_{i<j=1}^{2k+2}p_{i}\wed p_{j}+\sum_{i<\cT_{k}}p_{i}\wed\delta p_{l}+\sum_{\cT_{k}<\cT_{k}}\delta p_{l}\wed\delta p_{l'},\label{eq:VkOscill}\\
    V_{1}=&\delta(q_{1}+q_{2}+q_{3}+q_{4})\,\exp(-\imath\sum_{i<j=1}^{4}q_{i}\wed q_{j}).\label{eq:V1}
  \end{align}
  Let $\mathbf{l_{0}}$ the line joining a momentum $\mathbf{p_{i_{0}}},\,1\les i_{0}\les 2k+2$ to a momentum of $V_{1}$. By cyclicity of this one, one can assume that $l_{0}=(p_{i_{0}},\mathbf{q_{1}})$.
  We need to prove:
  \begin{align}
    V_{k}V_{1}=&\delta\Big(\sum_{i=1}^{4}q_{i}\Big)\delta\Big(\sum_{\substack{i=1\\i\neq i_{0}}}^{2k+2}p_{i}+\sum_{j=2}^{4}q_{j}+\sum_{l\in\cT_{k}}\delta p_{l}+\delta p_{l_{0}}\Big)\,\exp(-\imath\varphi_{k+1}),\label{eq:TRToProveDelta}\\
    \varphi_{k+1}=&\sum_{\substack{i<j=1\\i,j\neq i_{0}}}^{2k+2}p_{i}\wed p_{j}+\Big(\sum_{i<i_{0}}p_{i}-\sum_{i>i_{0}}p_{i}\Big)\wed\sum_{j=2}^{4}q_{j}+\sum_{i<j=2}^{4}q_{i}\wed q_{j}\nonumber\\
    &+\sum_{\substack{i<\cT_{k}\\i\neq i_{0}}}p_{i}\wed\delta p_{l}+\sum_{\cT_{k+1}<\cT_{k+1}}\delta p_{l}\wed\delta p_{l'}\nonumber\\
    &+\sum_{i<i_{0}}p_{i}\wed\delta p_{l_{0}}-\sum_{l\in\cT_{k},\,l>i_{0}}\delta p_{l}\wed\sum_{j=2}^{4}q_{j}\label{eq:Phik+1ToProve}
  \end{align}
  with $\mathbf{\cT_{k+1}}=\cT_{k}\cup\{l_{0}\}$. This would reproduce (\ref{eq:TreeReducOscill}).

  The statement concerning the delta function in (\ref{eq:TRToProveDelta}) is easily obtained. It only consists in the following distributional equality:
  \begin{align}
    \delta\Big(\sum_{i=1}^{2k+2}p_{i}+\sum_{l\in\cT_{k}}\delta p_{l}\Big)\delta\Big(\sum_{j=1}^{4}q_{j}\Big)=&\delta\Big(\sum_{\substack{i=1\\i\neq i_{0}}}^{2k+2}p_{i}+\sum_{j=2}^{4}q_{j}+\sum_{l\in\cT_{k}}\delta p_{l}+\delta p_{l_{0}}\Big)\delta\Big(\sum_{j=1}^{4}q_{j}\Big).
  \end{align}
  Let us now rewrite the oscillations. First of all note that thanks to the delta function in $V_{1}$, the oscillation of the Moyal vertex can be rewritten as $\exp\big(-\im\sum_{i<j=2}^{4}q_{i}\wed q_{j}\big)$. The complete oscillation in $V_{k}V_{1}$ is:
  \begin{align}
    \varphi_{k+1}=&\sum_{\substack{i<j=1\\i,j\neq i_{0}}}^{2k+2}p_{i}\wed p_{j}+\sum_{i<\cT_{k}}p_{i}\wed\delta p_{l}+\sum_{\cT_{k}<\cT_{k}}\delta p_{l}\wed\delta p_{l'}+\sum_{i<j=2}^{4}q_{i}\wed q_{j}+\delta\varphi,\label{eq:PhiK+1Start}\\
    \delta\varphi=&\sum_{i<i_{0}}p_{i}\wed p_{i_{0}}+p_{i_{0}}\wed\sum_{j>i_{0}}p_{j}=\Big(\sum_{i<i_{0}}p_{i}-\sum_{j>i_{0}}p_{j}\Big)\wed(-q_{1}+\delta p_{l_{0}})
    \intertext{where we used $p_{i_{0}}=-q_{1}+\delta p_{l_{0}}$. But $-q_{1}=q_{2}+q_{3}+q_{4}$ so that }
    \delta\varphi=&\Big(\sum_{i<i_{0}}p_{i}-\sum_{i>i_{0}}p_{i}\Big)\wed\sum_{j=2}^{4}q_{j}+\Big(\sum_{i<i_{0}}p_{i}-\sum_{j>i_{0}}p_{j}\Big)\wed\delta p_{l_{0}}.
  \end{align}
  We can now write
  \begin{align}
    \varphi_{k+1}=&\sum_{\substack{i<j=1\\i,j\neq i_{0}}}^{2k+2}p_{i}\wed p_{j}+\Big(\sum_{i<i_{0}}p_{i}-\sum_{i>i_{0}}p_{i}\Big)\wed\sum_{j=2}^{4}q_{j}+\sum_{i<j=2}^{4}q_{i}\wed q_{j}\nonumber\\
    &+\sum_{\substack{i<\cT_{k}\\i\neq i_{0}}}p_{i}\wed\delta p_{l}+\sum_{\cT_{k}<\cT_{k}}\delta p_{l}\wed\delta p_{l'}\nonumber\\
    &+\Big(\sum_{i<i_{0}}p_{i}-\sum_{j>i_{0}}p_{j}\Big)\wed\delta p_{l_{0}}+p_{i_{0}}\wed\sum_{l\in\cT_{k},\,i_{0}<l}\delta p_{l}.\label{eq:Phik+1Interm}
  \end{align}
  Comparing (\ref{eq:Phik+1Interm}) and (\ref{eq:Phik+1ToProve}) we see that it remains to prove
  \begin{align}
    p_{i_{0}}\wed\sum_{l\in\cT_{k},\,i_{0}<l}\delta p_{l}+\sum_{l\in\cT_{k},\,i_{0}>l}\delta p_{l}\wed p_{i_{0}}=&\Big(\sum_{l\in\cT_{k},\,l<i_{0}}\delta p_{l}-\sum_{l\in\cT_{k},\,l>i_{0}}\delta p_{l}\Big)\wed\Big(\delta p_{l_{0}}+\sum_{j=2}^{4}q_{j}\Big)\nonumber\\
    &+\frac 12 p_{l_{0}}\wed\delta p_{l_{0}}-\sum_{j=2}^{4}q_{j}\wed\delta p_{l_{0}}. 
  \end{align}
  We use $p_{i_{0}}=-q_{1}+\delta p_{l_{0}}=q_{2}+q_{3}+q_{4}+\delta p_{l_{0}}$ and get the equality if $\frac 12 p_{l_{0}}\wed\delta p_{l_{0}}-\sum_{j=2}^{4}q_{j}\wed\delta p_{l_{0}}=0$ which is true thanks to $q_{1}=\frac 12(\delta p_{L_{0}}-p_{l_{0}})$. This proves the lemma.
\end{proof}

\begin{lemma}[Rosette Factor]
  \label{lem:CompleteFilk}
  The rosette factor of a graph $G$ with $N(G)=N$ is given by
  \begin{align}
    &\delta\Big(\sum_{k=1}^{N}p_{k}+\sum_{l\in\cT\cup L}\delta p_{l}\Big)\,\exp(-\im\varphi)\text{ with }\varphi=\varphi_{E}+\varphi_{m}+\varphi_{\cap}+\varphi_{\lrtimes}+\varphi_{J},\label{eq:RosetteFactor}\\
    \nonumber\\
    \varphi_{E}=&\sum_{i<j=1}^{N}p_{i}\wed p_{j},\nonumber\\
    \nonumber\\
    \varphi_{m}=&\frac 12\sum_{\ell\in\cT\cup L}p_{\ell}\wed\delta p_{\ell}+\sum_{(\cT\cup L)\subset L}p_{\ell'}\wed\delta p_{\ell}+\frac 12\sum_{ L\ltimes L}(p_{\ell}\wed\delta p_{\ell'}+p_{\ell'}\wed\delta p_{\ell}),\nonumber\\
    \nonumber\\
    \varphi_{\cap}=&\sum_{ L\supset k}p_{\ell}\wed p_{k},\quad\varphi_{\lrtimes}=\frac 12\sum_{ L\ltimes L}p_{\ell}\wed p_{\ell'},\nonumber\\
    \nonumber\\
    \varphi_{J}=&\sum_{(\cT\cup L)<k}\delta p_{\ell}\wed p_{k}+\sum_{(\cT\cup L)>k}p_{k}\wed\delta p_{\ell}+\sum_{(\cT\cup L)<(\cT\cup L)}\delta p_{\ell}\wed\delta p_{\ell'}+\frac 12\sum_{ L\ltimes L}\delta p_{\ell}\wed\delta p_{\ell'}.\nonumber
  \end{align}
\end{lemma}
\begin{proof}
  Let us first fix an external momentum $p_{k}$. From lemma \ref{lem:TreeReduc} the linear term in $p_{k}$ is:
  \begin{align}
    &\Big(\sum_{i=1}^{k-1}p_{i}-\sum_{i=k+1}^{2n+2}p_{j}\Big)\wed p_{k}+\Big(\sum_{\cT<k}\delta p_{\ell}-\sum_{\cT>k}\delta p_{\ell}\Big)\wed p_{k}.
  \end{align}
  Let a line $\ell=(p_{\ell_{1}},p_{\ell_{2}})\in L$ such that $\ell<k$. Its contribution to this linear term is $(p_{\ell_{1}}+p_{\ell_{2}})\wed p_{k}=\delta p_{\ell}\wed p_{k}$. If $\ell>k$ we get $p_{k}\wed\delta p_{\ell}$. Let now $\ell\supset k$, we have $(p_{\ell_{1}}-p_{\ell_{2}})\wed p_{k}=p_{\ell}\wed p_{k}$. Then the linear term in the external momenta is:
  \begin{align}
    \sum_{(\cT\cup L)>k}p_{k}\wed\delta p_{\ell}+\sum_{(\cT\cup L)<k}\delta p_{\ell}\wed p_{k}+\sum_{ L\supset k}p_{\ell}\wed p_{k}.\label{eq:LinExt}    
  \end{align}
  Let us consider a line $\ell=(p_{\ell_{1}},p_{\ell_{2}})\in L$. The terms containing $p_{\ell_{1}}$ and $p_{\ell_{2}}$ are:
  \begin{align}
    &\sum_{i<\ell_{1}}p_{i}\wed p_{\ell_{1}}+\sum_{j>\ell_{1}}p_{\ell_{1}}\wed p_{j}+p_{\ell_{1}}\wed p_{\ell_{2}}+\sum_{i<\ell_{2}}p_{i}\wed p_{\ell_{2}}+\sum_{j>\ell_{2}}p_{\ell_{2}}\wed p_{j}\nonumber\\
    &+\sum_{\ell_{1}<\cT}p_{\ell_{1}}\wed\delta p_{\ell'}+\sum_{\ell_{1}>\cT}\delta p_{\ell'}\wed p_{\ell_{1}}+\sum_{\ell_{2}<\cT}p_{\ell_{2}}\wed\delta p_{\ell'}+\sum_{\ell_{2}>\cT}\delta p_{\ell'}\wed p_{\ell_{2}}.\label{eq:LoopsStart}\\
    =&\sum_{i<\ell_{1}}p_{i}\wed\delta p_{\ell}+\sum_{j>\ell_{2}}\delta p_{\ell}\wed p_{j}+\sum_{\ell_{1}<i<\ell_{2}}p_{\ell}\wed p_{i}+\sum_{\cT<\ell_{1}}\delta p_{\ell'}\wed\delta p_{\ell}+\sum_{\cT>\ell_{2}}\delta p_{\ell}\wed\delta p_{\ell'}\nonumber\\
    &+\sum_{\ell_{1}<\cT<\ell_{2}}p_{\ell}\wed\delta p_{\ell'}.\label{eq:LoopsStep1}
  \end{align}
  Let $\ell'=(p_{\ell'_{1}},p_{\ell'_{2}})\in L$ such that $\ell<\ell'$. From \eqref{eq:LoopsStart} one reads $(p_{\ell'_{1}}+p_{\ell'_{2}})\wed\delta p_{\ell}=\delta p_{\ell'}\wed\delta p_{\ell}$. If $\ell'\subset\ell$, one has $p_{\ell}\wed(p_{\ell'_{1}}+p_{\ell'_{2}})=p_{\ell}\wed\delta p_{\ell'}$. Finally if $\ell'\ltimes\ell$, one get $p_{\ell'_{1}}\wed\delta p_{\ell}+p_{\ell}\wed p_{\ell'_{2}}=\frac 12(p_{\ell'}+\delta p_{\ell'})\wed\delta p_{\ell}+p_{\ell}\wed\frac 12(\delta p_{\ell'}-p_{\ell'})$. We can now rewrite \eqref{eq:LoopsStep1} as:
  \begin{align}
    &\sum_{ L< L}\delta p_{\ell}\wed\delta p_{\ell'}++\sum_{ L\subset L}p_{\ell'}\wed\delta p_{\ell}+\frac 12\sum_{ L\ltimes L}(p_{\ell}\wed\delta p_{\ell'}+p_{\ell'}\wed\delta p_{\ell})+\frac 12\sum_{ L\ltimes L}p_{\ell}\wed p_{\ell'}\nonumber\\
    +&\frac 12\sum_{ L\ltimes L}\delta p_{\ell}\wed\delta p_{\ell'}+\sum_{\cT< L}\delta p_{\ell}\wed\delta p_{\ell'}+\sum_{\cT> L}\delta p_{\ell'}\wed\delta p_{\ell}+\sum_{\cT\cup\subset L}p_{\ell'}\wed\delta p_{\ell}\nonumber\\
    +&\frac 12\sum_{\ell\in L}p_{\ell}\wed\delta p_{\ell}.\label{eq:LoopsFinal}
  \end{align}
  Using lemma \ref{lem:TreeReduc} together with equations \eqref{eq:LinExt} and \eqref{eq:LoopsFinal}, one proves the lemma.
\end{proof}

\section{Power counting}
\label{sec:power-counting}

\subsection{The case \texorpdfstring{$\kappa$}{kappa}$=0$}
\label{sec:case-kappa=0}

As explained in the introduction, we give here a strong argument for the need of a new counterterm. Note also that the bound we obtain seems to be optimal in the sense that exact computations exhibit the same degree of divergence.

Remember also that one of our motivations is that it was noticed in \cite{Wang2008vn} that an harmonic oscillator term is not sufficient to make a scalar theory renormalisable on a \emph{degenerate} Moyal space. In this article the authors studied a $\phi^{\star 6}$ model on $\R^{3}_{\Theta}$ (see equation \eqref{eq:phi6Omega}) with the $x$-space representaion. They used the vertex delta functions to improve the usual commutative power counting $\omega=\frac 12(N-6+2v_{4})$ where $N$ is the number of external points and $v_{4}$ the number of four-valent vertices in the graph under consideration. For non-orientable graphs they got $\omega=\frac 12(N-2+2v_{4})$. This upper bound exhibits a logarithmic divergence for the (planar) non-orientable two-point graphs. The authors suggested that a possible solution to this problem may come from the use of the vertex oscillations. We now explain why we think that the solution should be looked for elsewhere.

To take the oscillations into account, a very powerful technique consists in using the matrix basis. On a degenerate Moyal space part of the coordinates commute and we must use a mixed representation. In the commutative directions, we choose the usual $x$- (or $p$-)space representation whereas we prefer the matrix basis in the \encv{} directions. On $\R^{3}_{\Theta}$, let us choose $\lsb x^{0},x^{i}\rsb=0,\,i=1,2$. Each field is expanded as
\begin{align}
  \phi(x)=&\sum_{m,n\in\Z}\phi_{m,n}(x^{0})f_{m,n}(x^{1},x^{2})\label{eq:FieldExpMatrix}
\end{align}
where the functions $f_{m,n}$ form a basis for the Schwartz-class functions. Then we get a representation of the model which is partly commutative and local (in the $x^{0}$-direction) and partly in the matrix basis. We can now apply the method developped in \cite{Rivasseau2005bh} to get an improved power counting namely
\begin{align}
  \omega=&\frac 12(N-6+8g+4(b-1)+2v_{4})\label{eq:PowCountphi6}
\end{align}
where $g$ is the genus of the graph and $b$ its number of broken faces.

The conclusion is that, whereas we took the oscillations into account, there still remains potentially log.\@ divergent two-point graphs with two broken faces. Hence the addition of an harmonic potential in the non-commutative directions does not imply renormalisability on a degenerate Moyal space. 

Back to our model \eqref{eq:ActionMinim}, it is clear that we can easily apply the same kind of mixed representation and get
\begin{lemma}
  \label{lem:PowCountPhi4Matrix}
  Let $G$ be a Feynman graph corresponding to the model \eqref{eq:ActionMinim} at $\kappa =0$. Its degree of convergence obeys the following bound:
  \begin{align}
    \omega(G)\ges N-4+4g+2(b-1).\label{eq:PowCount}
  \end{align}
\end{lemma}
\begin{proof}
  The quadratic parts corresponding to the commutative and \encv{} directions commute with each other. Therefore the Schwinger representation of the corresponding propagators factorize. Then to prove the lemma it is enough to apply the standard bounds in the commutative directions and the method developped in \cite{Rivasseau2005bh} in the \encv{} directions.
\end{proof}

Recall that on the fully \encv{} $\R^{4}_{\Theta}$, the power counting is $\omega\ges N-4+8g+4(b-1)$. The $N-4$ part is the usual power counting of the commutative $\phi^{4}$ model whereas the rest has a purely \encv{} origin. On the degenerate four-dimensional Moyal space, only half of the directions are \encv{} so that we gain only half of $8g+4(b-1)$ and get \eqref{eq:PowCount}. The consequence is that the planar two-point graphs with two broken faces ($N=b=2,\,g=0$) diverge logarithmically. They must be renormalised.
\begin{rem}
  One could ask if the addition of an harmonic potential also in the commutative directions could solve the problem. Unfortunately one can easily convince oneself (for example in $p$-space) that such an infrared modification of a commutative and local model doesn't change anything to the power counting.
\end{rem}

\subsection{The case \texorpdfstring{$\kappa\neq}{kappa non} 0$}
\label{sec:case-kappaneq-0}

In this subsection we are going to compare the power countings of graphs with and without insertions. To this aim, we recall briefly how to get the bound $\omega(G_{0})\ges N(G_{0})-4$ on the degree of convergence of a graph $G_{0}$ without $\kappa$-insertions in momentum space. We first have to perform the so-called momentum routing. This is the optimal way of using the delta functions attached to each Moyal vertex. For this we must choose a spanning rooted tree in $G_{0}$. Then we associate to this tree a set of branch\footnote{Given a spanning rooted tree $\cT(G)$ and a line $l\in\cT$, the branch $b(l)$ is the set of vertices $v$ such that $l$ is on the unique path in the tree between $v$ and the root of $\cT$.} delta functions which allow to solve $v(G_{0})-1$ momenta both in the commutative and \encv{} directions \cite{RenNCGN05}.

In a slice $i$, a propagator is bounded by
\begin{align}
  \Ch^{i}(p,\kp;p,\kq)\les K\,e^{-M^{-2i}p^{2}}e^{-M^{2i}(\kp+\kq)^{2}-M^{-2i}(\kp-\kq)^{2}}.\label{eq:BoundPropa}
\end{align}
For each line $l$ of $G_{0}$, the integration over $\kp_{l}+\kq_{l}$ gives a factor $M^{-2i_{l}}$. We recover the power counting factor of $\frac{1}{p^{2}}$ in four dimensions. Then if $l$ is a loop line, the integrations over $p_{l}$ and $\kp_{l}-\kq_{l}$ deliver together $M^{4i_{l}}$. If $l$ is a tree line, these integrations are made with a delta function and bring $\cO(1)$. We get the bound
\begin{align}
  |A_{G_{0}}|\les&K^{n}\prod_{l\in E(G)}M^{-2i_{l}}\prod_{l\in L(G)}M^{4i_{l}}\les K^{n}\prod_{i,k}M^{-\omega(G^{i}_{k})},\,\omega=N-4.
\end{align}

We now turn to the computation of an upper bound on the amplitude of a graph $G$ with $\kappa$-insertions. The graph $G$ is equipped with a scale attribution $\mu(G)$ which assigns an integer to each edge in $E(G)$. To get the power counting of such a graph we need to perform the momentum routing and pick up a tree $\cT_{G}$. Note that each generalised line is considered as one single line so that the tree $\cT_{G}$ is composed of both simple and generalised lines.

Let us focus on a generalised line $\ell$ between two Moyal vertices. It is made of $n(\ell)$ insertions and so $n+1$ edges $\ell_{k},\,k\in\lnat 1,n+1\rnat$. The corresponding analytical expression is:
\begin{align}
  A_{\mathbf{\ell}}\defi&\kappa^{2n}\Ch^{i_{\ell_{1}}}(p,\kp;p,0)\Big(\prod_{k=2}^{n}\Ch^{i_{\ell_{k}}}(p,0;p,0)\Big)\Ch^{i_{\ell_{n+1}}}(p,0;p,\kq).\label{eq:GenLineExpr1}
\end{align}
Thanks to the bound \eqref{eq:BoundPropa}, we have
\begin{align}
  A_{\mathbf{\ell}}\les&\kappa^{2n}K^{n+1}\,e^{-(\sum_{k=1}^{n+1}M^{-2i_{\ell_{k}}})p^{2}}e^{-M^{2i_{\ell_{1}}}\kp^{2}-M^{2i_{\ell_{n+1}}}\kq^{2}}.\label{eq:BoundAl}
\end{align}
Let $\mathbf{i_{\text{\textbf{m}}}}\defi\min_{k\in\lnat 1,n+1\rnat}i_{\ell_{k}}$, $\mathbf{i_{1}}\defi\max\{i_{\ell_{1}},i_{\ell_{n+1}}\}$ and $\mathbf{i_{2}}\defi\min\{i_{\ell_{1}},i_{\ell_{n+1}}\}$. Then exchanging $\kp$ and $\kq$ if necessary, $A_{\ell}$ is bounded by
\begin{align}
  A_{\mathbf{\ell}}(p;\kp,\kq)\les&\kappa^{2n}K^{n+1}\,e^{-(n+1)M^{-2i_{\text{m}}}p^{2}}e^{-M^{2i_{1}}\kp^{2}-M^{2i_{2}}\kq^{2}}.\label{eq:BoundAl2}
\end{align}
In the following, we consider that a generalised line $\ell$ is a line of scale $i_{\ell}\defi i_{\text{m}}(\ell)$. This will be important in the choice of an optimised tree.

\begin{rem}[No Moyal vertex]
  The bound (\ref{eq:BoundAl2}) does ot depend on the scales $i_{\ell_{2}},\dots,i_{\ell_{n}}$. This reflects the fact that the corresponding subgraphs are logarithmically divergent. These graphs are made of propagators linked together by $\kappa$-insertions but do not contain any Moyal vertex. Let $G$ be such a subgraph with $n$ propagators. The corresponding analytical expression is:
  \begin{align}
    A_{G}=&\int_{\R^{2}} d^{2}p\,\phih(p,0)\phih(-p,0)\prod_{k=2}^{n}\Ch^{i_{\ell_{k}}}(p,0;p,0).
  \end{align}
  To renormalise such a graph, we expand the propagators around $p=0$:
  \begin{align}
    \Ch^{i_{\ell}}(p,0;p,0)=&\Ch^{i_{\ell}}(0,0;0,0)+\int_{0}^{1}ds\,p\cdot\nabla\Ch^{i_{\ell}}(sp,0;sp,0)
  \end{align}
  The graph $G$ being only log.\@ divergent, only the zeroth order term is divergent. It contributes to the renormalisation of $\kappa^{2}$.

In the following of this article, we will not anymore make any reference to these particular subgraphs keeping nevertheless in mind that they are log.\@ divergent and can be renormalised by a change of $\kappa^{2}$.
\end{rem}

\begin{figure}[htb]
  \centering
  \subfloat[Admissible line]{{\label{fig:AdmLine}}\includegraphics[scale=1]{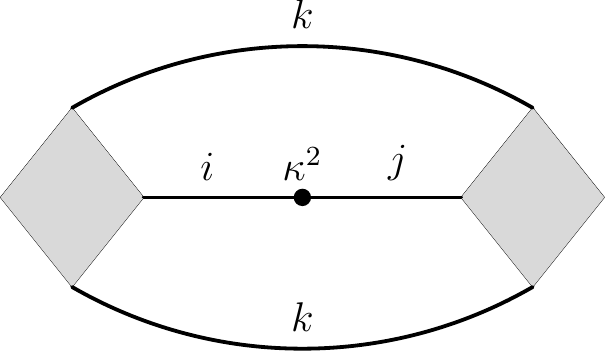}}\hspace{2cm}\quad
  \subfloat[Non-admissible line]{\label{fig:NonAdmLine}\includegraphics[scale=1]{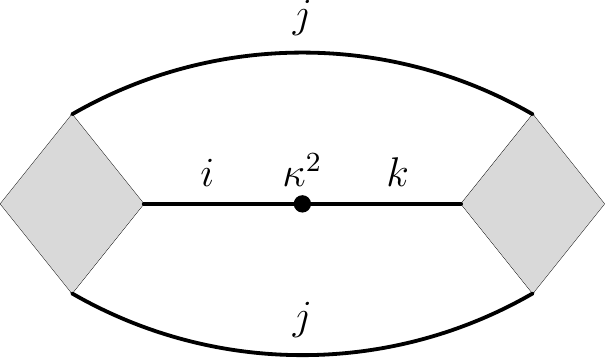}}
  \caption{Scale attribution and admissibility, $i>j>k$}
  \label{fig:ScaleAttAdm}
\end{figure}

Before we state the power counting lemma, we need to give a few definitions:
\begin{defn}[Bridge]
  Let $G$ be a graph and $l\in E(G)$. The line $l$ is a \emph{bridge} if deleting $l$ increases the number of connected components of $G$.
\end{defn}
\begin{defn}[Admissible generalised line]
  Let $G^{\mu}$ be a graph with scale attribution $\mu$ and $l\in E_{\kappa}(G)$. There exists a unique $k(l)\in\N$ such that $l\in E_{\kappa}(G^{i_{l}}_{k})$. The line $l$ is said \emph{admissible} if it is a bridge in $G^{i_{l}}_{k}$.
\end{defn}
The admissibility of a generalised line depends both on the location of the line in the graph and on the scale attribution as shown in figure \ref{fig:ScaleAttAdm}.
\begin{defn}[Tree-like graph]
  Let $G$ be a graph. It is said \emph{tree-like} if for all $l\in E_{\kappa}(G)$, $l$ is a bridge (independently of the scale attribution). By convention, a graph without insertion is tree-like.
\end{defn}
A tree-like graph is then a tree of generalised lines the nodes of which are graphs with only simple lines and Moyal vertices.
\begin{lemma}[Power counting]
  \label{lem:PowCount}
  Let $G$ be a Feynman graph of the model \eqref{eq:ActionMinim}. Let $\mu(G)$ be a scale attribution. Then we have:
  \begin{subequations}
    \begin{itemize}
    \item if $G$ is not tree-like, the amplitude $A_{G}^{\mu}$ is
      bounded by
      \begin{align}
        |A_{G}^{\mu}|\les&K^{v(G)}\prod_{i,k}M^{-\omega(G^{i}_{k})},\
        \omega\ges N+4n_{\kappa}\label{eq:PowCountNotTreeLike}
      \end{align}
      for the connected components which are not tree-like.
    \item If $G$ is tree-like, for all connected component
      $G^{i}_{k}$, its degree of convergence obeys:
      \begin{itemize}
      \item if $G^{i}_{k}$ is non planar
        \begin{align}
          \omega\ges&N\label{eq:PowCountTreeLikeg}
        \end{align}
      \item if $g(G^{i}_{k})=0$, $b(G^{i}_{k})\ges 2$
        \begin{align}
          \omega\ges&N-2,\label{eq:PowCountTreeLikeb}
        \end{align}
      \item if $G^{i}_{k}$ is planar regular and
        $E_{\kappa}(G^{i}_{k})\neq\emptyset$
        \begin{align}
          \omega\ges&N-2\label{eq:PowCountTreeLikeRegWith}
        \end{align}
      \item if $G^{i}_{k}$ is planar regular and
        $E_{\kappa}(G^{i}_{k})=\emptyset$
        \begin{align}
          \omega\ges&%
          \begin{cases}
            N-4+2N_{\kappa}&\text{if $N_{\kappa}<N$}\\
            N-4+2(N_{\kappa}-1)&\text{if $N_{\kappa}=N$.}
          \end{cases}\label{eq:PowCountTreeLikeRegOhne}
        \end{align}
      \end{itemize}
    \end{itemize}
  \end{subequations}
\end{lemma}
This lemma proves that the only graphs (and subgraphs) which need to be renormalised are tree-like. Moreover they are either planar and regular: in that case, if they have four external points, they don't have any insertion. If they are two-point graphs, they have zero, one or two external insertions and possibly internal ones also. Or they are planar with two broken faces.

\subsection{Proof ($1/2$): truncated diagrams}
\label{sec:proof-first-part}

In this section we prove all the bounds of lemma \ref{lem:PowCount} except the improvement related to the number of external insertions $N_{\kappa}$. This is postponed to subsection \ref{sec:proof-second-part}.\\

Let $G$ be a Feynman graph. Its amplitude is given by:
\begin{align}
  A_{G}^{\mu}(\wp_{1},\dots,\wp_{N})=&\int_{\R^{8e}}\prod_{l\in E(G)}d^{4}p_{l_{1}}d^{4}p_{l_{2}}\,\Ch^{i_{l}}(p_{l_{1}},p_{l_{2}})\prod_{v\in V(G)}\delta_{v}\,e^{-\imath\hat{\varphi}}\label{eq:AG}
\end{align}
To get the bound (\ref{eq:PowCountNotTreeLike}), we start by bounding the oscillations by one. Then we perform a momentum routing. To this aim we choose an optimised spanning rooted tree\footnote{Here ``optimised'' means that the tree is a subtree in each connected component.} and trade the vertex delta functions for an equivalent set of branch delta functions. This allows to solve one momentum per tree line: at each vertex (except the root) the delta function solves the unique momentum hooked to this vertex which is on the path in the tree between the vertex and the root. For a simple line, let us call the combination $\kp+\kq$ a \emph{\textbf{short}} variable. The momentum routing replaces the solved tree momenta by $p_{L}+p_{Ex}+\delta p$ where $p_{L}$ (resp.\@ $p_{Ex}$, $\delta p$) is a linear combination of loop momenta of \emph{simple} line (resp.\@ external momenta, short variables or momenta of generalised lines). As a result we have to integrate
\begin{itemize}
\item in the commutative directions, over one momentum per loop line (thanks to the conservation of momentum along the lines)
\item and in the \encv{} directions, over one momentum per tree line and two momenta per loop line.
\end{itemize}
Moreover there remains a global delta function which ensures the exact conservation of the external momenta in the commutative directions and an approximate conservation in the \encv{} directions (see \cite{RenNCGN05} for details about an equivalent \emph{position} routing).\\

Let $l\in\cT\cap E_{0}$. In the commutative directions, its corresponding momentum has been solved thanks to the delta function $\delta_{b(l)}$. In the \encv{} directions, this delta function allows to solve $\kp_{l}-\kq_{l}$ (see equation (\ref{eq:BoundPropa})). We still have to integrate over $\kp_{l}+\kq_{l}$ which delivers a factor bounded by $M^{-2i_{l}}$.

Let $\ell\in\cT\cap E_{\kappa}$. We use the delta function corresponding to the branch $b(\ell)$ to integrate over $p$ and either $\kp$ or $\kq$. The result is bounded by:
\begin{align}
  \int_{\R^{6}}d^{2}p\,d^{2}\kp\,d^{2}\kq\,A_{\ell}(p;\kp,\kq)\delta_{b(\ell)}\les&K M^{-2(i_{2}-i_{\text{m}})}M^{-2i_{\text{m}}}.\label{eq:AlTree}
\end{align}

Let $l\in L\cap E_{0}$. We integrate over $p,\kp$ and $\kq$. Thanks to the bound (\ref{eq:BoundPropa}), the result is bounded by $M^{-2i_{l}}M^{4i_{l}}$.

Let $\ell\in L\cap E_{\kappa}$. We have to integrate $A_{\ell}$ over $p,\kp$ and $\kq$. The result is bounded by:
\begin{align}
  \int_{\R^{6}}d^{2}p\,d^{2}\kp\,d^{2}\kq\,A_{\ell}(p;\kp,\kq)\les&K M^{-2(i_{1}-i_{2})}M^{-4(i_{2}-i_{\text{m}})}M^{-2i_{\text{m}}}.\label{eq:AlLoop}
\end{align}
As a consequence, we have
\begin{align}
  |A^{\mu}_{G}|\les K^{v(G)}\prod_{l\in E}M^{-2i_{l}}\prod_{l\in L\cap E_{0}}M^{4i_{l}}\prod_{l\in E_{\kappa}}M^{-2(i_{2}-i_{\text{m}})}\prod_{l\in L\cap E_{\kappa}}M^{-2(i_{1}-i_{2})}.\label{eq:IntermWithExt}
\end{align}
The last two products are clearly related to the external insertions and contribute to the improvement of the power counting by the factor $2N_{\kappa}$. In subsection \ref{sec:proof-second-part}, we will explain how to improve these factors to reproduce completely the bounds of lemma \ref{lem:PowCount}. Until there we just bound these products by one. Then we have
\begin{align}
  |A^{\mu}_{G}|\les K^{v(G)}\prod_{l\in E}M^{-2i_{l}}\prod_{l\in L\cap E_{0}}M^{4i_{l}}\les K^{v(G)}\prod_{i,k}M^{-\omega(G^{i}_{k})},\text{ with $\omega=N-4+4n_{\kappa}$.}\label{eq:IntermBound1}
\end{align}
This proves that if a subgraph $G_{k}^{i}$ is divergent, $n_{\kappa}(G_{k}^{i})=0$ and all its generalised lines are in the tree. This means that for all $l\in E_{\kappa}(G_{k}^{i})$ and for \emph{any} choice of an \emph{optimised} tree in $G_{k}^{i}$, $l$ is in the tree. This implies that $l$ is a bridge in a $G^{i_{l}}_{k'}$. In other words if a connected component is divergent, all its generalised lines are admissible. Nevertheless this doesn't prove yet the bound (\ref{eq:PowCountNotTreeLike}) because an admissible line $l$ is only a bridge in a $G^{i_{l}}_{k}$ but not necessarily in the full graph $G$, as shown in the figure \ref{fig:AdmLine}. We have to improve our bound.

Let us consider a connected component $G^{i}_{k}$ and an admissible generalised line $l\in E_{\kappa}(G_{k}^{i})$ which is not a bridge in $G_{k}^{i}$. This line belongs to the tree $\cT(G_{k}^{i})$ so that we use the delta function $\delta_{b(l)}$ to integrate over $\kp_{l}$ or $\kq_{l}$. In the propagator $\Ch^{i_{l}}$, $\kp_{l}$ (say) is replaced by $p_{L}+p_{Ex}+\delta p$. Usually, we bound $\Ch^{i_{l}}$ by $M^{-2i_{l}}$ but if $p_{L}\neq 0$, we can use it to integrate over a loop momentum to get $M^{-2i_{l}}\les M^{-2i}=M^{2i}M^{-4i}$. The gain is $M^{-4i}$ and makes $G_{k}^{i}$ convergent. Thus for any connected component $G_{k}^{i}$ and any line $l\in E_{\kappa}(G_{k}^{i})$ with $i_{l}\ges i$, $G_{k}^{i}$ divergent implies that $l$ is a bridge in $G_{k}^{i}$. All the generalised lines have to be bridges in $G=G^{0}$ and so $G$ has to be tree-like to be divergent. If not, the improvement factor $M^{4}$ plus the bound (\ref{eq:IntermBound1}) give the equation (\ref{eq:PowCountNotTreeLike}).\\

We have proven that if a graph is divergent, all its connected component are tree-like which is equivalent to $G$ itself being tree-like. So let us consider such a graph and prove the bounds (\ref{eq:PowCountTreeLikeg}~-~\ref{eq:PowCountTreeLikeRegWith}).

We start with the bound (\ref{eq:IntermBound1}) with $n_{\kappa}=0$ (because $G$ is tree-like). We are going to improve it thanks to the oscillations of $A_{G}$. In \cite{xphi4-05}, one of us contributed to proving, in $x$-space, that non-planar graphs are convergent. We just use here the same method but in momentum space: if two line $l$ and $l'$ cross each other, there exists an oscillation of the type $p_{l}\wed p_{l'}$ (see lemma \ref{lem:CompleteFilk}). Note that the presence of generalised lines does not alter that result because they are tree lines and only loop lines may cross each other. We use such an oscillation to integrate over $p_{l'}$ say. The gain with respect to the bound (\ref{eq:IntermBound1}) is $M^{-2(i_{l}+i_{l'})}\les M^{-4\min\{i_{l},i_{l'}\}}$. It leads to the bound (\ref{eq:PowCountTreeLikeg}).\\

Let us now consider a planar connected component but with at least two broken faces. Then there exists an oscillation of the type $p_{l}\wed p_{e}$ where $p_{e}$ is an external momentum of the subgraph. Once more this oscillation allows to integrate over $p_{e}$. The gain is $M^{-2i_{l}}$ and gives the bound (\ref{eq:PowCountTreeLikeb}).\\

Let $G$ be a planar regular tree-like graph. The bound (\ref{eq:IntermBound1}) for $n_{\kappa}=0$ gives already $\omega\ges N-4$. To improve it and get (\ref{eq:PowCountTreeLikeRegWith}), we must prove the following: given a tree-like graph $G$ and a generalised line $l$, $l$ is necessarily on the path in the tree between two external points.

\begin{figure}[htb]
  \centering
  \includegraphics[scale=1]{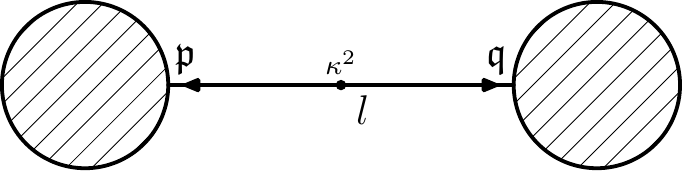}
  \caption{A tree-like graph}
  \label{fig:Bridge}
\end{figure}

\noindent
The graph being tree-like, $l$ is a bridge. $G$ can consequently be depicted as in figure \ref{fig:Bridge} where the two blobs represent any tree-like graphs. To each of these graphs, an \emph{odd} number of external points are hooked. In particular each of them contains at least one external point.

\noindent
Thanks to the momentum routing, the momentum $\kq$ (say) equals minus the sum of the momenta entering the right blob. Then we can use the propagator of the line $l$ to integrate over one of these external momenta, in the \encv{} directions. From the bound (\ref{eq:BoundAl2}), the result is bounded by $M^{-2i_{l}}$. This makes the improvement from (\ref{eq:IntermBound1}) to (\ref{eq:PowCountTreeLikeRegWith}).\\

There now remains to prove how appear the factors $N_{\kappa}$ in (\ref{eq:PowCountTreeLikeRegOhne}). This is the subject of the next section.

\subsection{Proof $(2/2)$: external insertions}
\label{sec:proof-second-part}

The basic mechanism which implies an improvement of the power counting thanks to the external insertions is the following.

Let us consider a graph $G$ with $N_{\kappa}$ external insertions and the lowest scale of which is $j$. It has so $N_{\kappa}$ external legs which correspond to $\Ch^{i}(p,\kp;p,0)$. Thanks to the equation (\ref{eq:BoundAl2}), the integration over $\kp$ gives a factor $M^{-2i}\les M^{-2j}$. If $N_{\kappa}<N$, the graph has $N-N_{\kappa}$ external momenta and $N_{\kappa}$ external insertions. We use the global delta function of $G$ to solve one external momentum in function of the others. Then each external insertion delivers a factor $M^{-2i}$. If $N_{\kappa}=N$, the global delta function solves the momentum of one of these insertions. This gives the degree of convergence in (\ref{eq:PowCountTreeLikeRegOhne}).\\

Nevertheless we still have to prove that the procedure which lead to the bound (\ref{eq:IntermBound1}) reproduces this improvement in all the connected components. Of course this is related to the last two products in equation (\ref{eq:IntermWithExt}).

\noindent
Let us consider a connected component $G^{i}_{k}$ with $N_{\kappa}$ external insertions. These insertions correspond to generalised lines at lower scales. If these lines are loop lines, the bound \eqref{eq:AlLoop} gives a factor $M^{-2(i_{1}-i_{2})}M^{-4(i_{2}-i_{\text{m}})}$ which is precisely the gain of two powers per external insertion. But for the generalised lines which are in the tree, the argument is subtler. Each such line bears two momenta, one of them being solved by the momentum routing. This is the ``highest'' of the two in the tree. The momentum routing solves then at most one such momentum among the momenta corresponding to the external insertions. If $N_{\kappa}=N$, this corresponds to the fact that the global delta function $\delta_{G^{i}_{k}}$ solves this momentum and that we cannot get a better improvement than $2(N-1)$. If $N_{\kappa}<N$, it could very well happen that an external insertion is on the unique path between $G^{i}_{k}$ and the root of $\cT(G)$. In this case, we would not get a gain of $2N_{\kappa}$. But we can use the propagator of that external insertion to integrate over an external momentum (which is not another external insertion). The result is $M^{-2i}=M^{-2(i-i_{\text{m}})}M^{-2i_{\text{m}}}$. This reproduces the bound (\ref{eq:PowCountTreeLikeRegOhne}) and is also compatible with (\ref{eq:PowCountTreeLikeRegWith}). This ends the proof of lemma \ref{lem:PowCount}.

% \paragraph{Locally rooted momentum routing} Given a graph $G$ and a scale attribution $\mu$, we pick up an optimised spanning tree $\cT$. Being optimised the tree is subtree in all the connected components. We can now proceed in two different ways. We can either choose one (global) root for the tree $\cT$ or one root per connected component. In the first case, each branch delta function $\delta_{b(l)}$ contain only one tree momentum namely the one which flows on the line $l$. In the local case, given a connected component $G^{i}_{k}$, each branch delta function $\delta_{b(l)}$ for $l\in\cT(G^{i}_{k})$ only depends on one tree momentum of scale higher than $i$ but possibly on several of lower scales.\\

% Now given a connected component $G^{i}_{k}$ and a tree $\cT(G^{i}_{k})$, we can always choose a root such that if $N_{\kappa}(G^{i}_{k})<N(G^{i}_{k})$, the global delta function $\delta_{G^{i}_{k}}$ solves an external momentum and not an external insertion.

\section{Renormalisation}
\label{sec:renormalisation}

Thanks to the power counting lemma \ref{lem:PowCount} we know which types of graphs are divergent. In this section, we prove that the divergent parts of these graphs reproduce the five terms of the lagrangean (\ref{eq:ActionMinim}).

\subsection{The four-point function}
\label{sec:four-point-function}

The only divergent four-point graphs are planar regular and contain no $\kappa$-insertion (neither internal nor external). The ``Moyality'' of the corresponding Feynman amplitudes has already been proven in \cite{xphi4-05} in the case of a non-degenerate Moyal space using the $x$-space representation. The only differences here are that we use the momentum space representation and that our \encv{} space is half commutative. Nevertheless, our proof would be so close to the one in \cite{xphi4-05} that we do not feel the need to reproduce it here.

\subsection{The two-point function}
\label{sec:two-point-function}

\subsubsection{The planar regular case}
\label{sec:planar-regular-case}

Let $G$ be a planar regular two-point graph. We distinguish mainly three different cases: 
\begin{enumerate}
\item $E_{\kappa}(G)=\emptyset \text{ and } N_{\kappa}(G)=0$,
\item $E_{\kappa}(G)\neq\emptyset \text{ and } N_{\kappa}(G)=0$ and\label{item:1}
\item $N_{\kappa}(G)\neq 0$.\label{item:2}
\end{enumerate}

\paragraph{No insertion at all}
\label{sec:no-insertion-at}

As in the case of the four-point function, there is no major difference between our degenerate model and the case treated in \cite{xphi4-05}. The two-point graphs contribute to the flow of the mass, wave-functions\footnote{The coefficients in front of the Laplacean in the commutative and \encv{} directions renormalise separately.} and oscillator frequency $\Omega$.

\paragraph{With internal insertions}
\label{sec:with-intern-insert}

Let $G$ be a connected planar regular tree-like graph with $E_{\kappa}\neq\emptyset$ and $N_{\kappa}=0$. Let $i$ be the lowest of its scales. We now prove that its divergent part renormalises $\kappa^{2}$.

For all ligne $l\in E(G)$, let $p_{L}^{l}$ be a linear combination of loop momenta in the commutative directions. Let $\kp_{L}^{l}$ (resp.\@ $\delta\kp_{l}$) be a linear combination of loop momenta (resp.\@ short variables and momenta of generalised lines) in the \encv{} directions. Finally let $\delta\kp$ be the sum of all the short variables of $G$ plus the sum of all the momenta of the generalised lines. The amplitude of $G$ integrated over external fields and after a momentum routing is given by:
\begin{align}
  A^{\mu}_{G}\fide&\int_{\R^{6}}d^{2}p\,d^{2}\kp\,d^{2}\kq\,\phih(p,\kp)\phih(-p,\kq)\cA(p,\kp,\kq)\label{eq:DefcA}\\
  =&\int_{\R^{4+2(v-1)+6n}}d^{2}p\,d^{2}\kp\,\phih(p,\kp)\phih(-p,-\kp-\delta\kp)\,e^{\imath\hat{\varphi}}\prod_{l\in L(G)}d^{2}p_{l}\,d^{2}\kp_{l}\,d^{2}\kq_{l}\,\Ch^{i_{l}}(p_{l},\kp_{l};p_{l},\kq_{l})\nonumber\\
  &\prod_{l\in\cT\cap E_{0}}d^{2}\kq_{l}\,\Ch^{i_{l}}(p+p^{l}_{L},\kq_{l}+\kp+\kp^{l}_{L}+\delta\kp_{l};p+p^{l}_{L},\kq_{l})\prod_{l\in\cT\cap E_{\kappa}}d^{2}\kq_{l}\,\Ch^{i_{l}}(p,\kp+\delta\kp_{l};p,\kq_{l}).\nonumber
\end{align}
The oscillation is of the type $\hat{\varphi}=\kp\wed\delta\kp$.

Note that after the momentum routing, there is no loop momenta in the generalised lines. This is because all such lines are bridges. This allows to bound the external momentum $\kp$ by $|\kp|\les M^{-i}$. We then perform a Taylor expansion of the external fields around $\kp=0$. In the commutative directions, as usual, we expand the tree propagators around $p=0$. Thanks to the power counting lemma, we know that such an amplitude is only log.\@ divergent. As a consequence, only the zeroth order term of these expansions is divergent.
\begin{align}
  A^{\mu}_{G}=&\int_{\R^{2}}d^{2}p\,\phih(p,0)\phih(-p,0)\int_{\R^{2+2(v-1)+6n}}d^{2}\kp\,e^{\imath\hat{\varphi}}\prod_{l\in L(G)}d^{2}p_{l}\,d^{2}\kp_{l}\,d^{2}\kq_{l}\,\Ch^{i_{l}}(p_{l},\kp_{l};p_{l},\kq_{l})\nonumber\\
  &\prod_{l\in\cT\cap E_{0}}d^{2}\kq_{l}\,\Ch^{i_{l}}(p^{l}_{L},\kq_{l}+\kp+\kp^{l}_{L}+\delta\kp_{l};+p^{l}_{L},\kq_{l})\prod_{l\in\cT\cap E_{\kappa}}d^{2}\kq_{l}\,\Ch^{i_{l}}(0,\kp+\delta\kp_{l};0,\kq_{l})\nonumber\\
  &+\text{ convergent contributions.}
\end{align}
The planar regular tree-like graphs with internal insertions contribute to the renormalisation of $\kappa^{2}$.

\paragraph{With external insertions}
\label{sec:with-extern-insert}

Let $G$ be a connected planar regular tree-like graph with $N_{\kappa}\neq 0$. Let $i$ be the lowest of its scales. We now prove that its divergent part renormalises $\kappa^{2}$. Let us first consider that $G$ has only one external insertion. In this case, its amplitude is
\begin{align}
  A^{\mu}_{G}=&\int_{\R^{6}}d^{2}p\,d^{2}\kp\,d^{2}\kq\,\phih(p,0)\Ch(p,0;p,\kp)\cA(p,\kp,\kq)\phih(-p,\kq)
\end{align}
where $\cA$ is defined by the equation (\ref{eq:DefcA}). In contrast with the previous case, even if $E_{\kappa}=\emptyset$, the external momentum $\kp$ (and consequently $\kq$) is bounded by $M^{-i}$. This is due to the external insertion. Then we can safely expand the external field around $\kq=0$, and $\cA$ and the external propagator around $p=0$. This leads to
\begin{align}
  A^{\mu}_{G}=&\int_{\R^{2}}d^{2}p\,\phih(p,0)\phih(-p,0)\int_{\R^{4}}d^{2}\kp\,d^{2}\kq\,\Ch(0,0;0,\kp)\cA(0,\kp,\kq)\text{ $+$ convergent terms.}
\end{align}
If $N_{\kappa}(G)=2$, the amplitude is
\begin{align}
  A^{\mu}_{G}=&\int_{\R^{6}}d^{2}p\,d^{2}\kp\,d^{2}\kq\,\phih(p,0)\Ch(p,0;p,\kp)\cA(p,\kp,\kq)\Ch(-p,\kq;-p,0)\phih(-p,0).
\end{align}
We expand $\cA$ and the external propagators around $p=0$. The divergent part of this expansion renormalises $\kappa^{2}$ too.

\subsubsection{The planar irregular case}
\label{sec:plan-irreg-case}

Let $G$ be a planar irregular ($b(G)=2$) two-point graph. Let us first treat the case of a graph without any insertion. Its amplitude would be
\begin{align}
  A^{\mu}_{G}=&\int_{\R^{4+2(v-1)+6n}}d^{2}p\,d^{2}\kp\,\phih(p,\kp)\phih(-p,-\kp-\delta\kp)\,e^{\imath\hat{\varphi}}\prod_{l\in L(G)}d^{2}p_{l}\,d^{2}\kp_{l}\,d^{2}\kq_{l}\,\Ch^{i_{l}}(p_{l},\kp_{l};p_{l},\kq_{l})\nonumber\\
  &\prod_{l\in\cT}d^{2}\kq_{l}\,\Ch^{i_{l}}(p+p^{l}_{L},\kq_{l}+\kp+\kp^{l}_{L}+\delta\kp_{l};p+p^{l}_{L},\kq_{l})\nonumber
\end{align}
where the oscillation is of the type $\hat{\varphi}=\kp\wed(\delta\kp +\kp_{L})$ (see lemma \ref{lem:CompleteFilk}). The oscillation between the external momentum $\kp$ and some loop momenta $\kp_{L}$ allows to prove that $\kp$ is actually bounded by $M^{-i}$ (see \cite{RenNCGN05} for the details). We can then perform the same types of expansions as before to get a renormalisation of $\kappa^{2}$.

If $G$ contains internal or external insertions, it should now be clear that its amplitude contributes also to the flow of $\kappa^{2}$. Indeed, whatever the reason, if one can control the size of the (\encv{}) external momentum, one can expand the external fields around $0$ as was done in the preceding cases.

\section{Conclusion}
\label{sec:conclusion}

Motivated by the work of Wan and Wang \cite{Wang2008vn} and by the possibility of defining a renormalisable model on \encv{} Minkowski space, we addressed here the problem of the renormalisability of a self-interacting quantum field on a degenerate Moyal space. On such a space, part of the coordinates are commutative. Contrary to the non-commutative $\Phi^{\star 4}_{4}$ model \cite{GrWu04-3}, the harmonic oscillator term is not sufficient to make the model renormalisable. We proved that the model contains indeed additionnal divergencies of the type $(\Tr\phi)^{2}$. By adding such a counterterm, we defined a renormalisable model (see \eqref{eq:ActionMinim}).

The interest for such a study is twofold. On one side, the appearence of such counterterms (of the type ``product of traces'') is quite natural on \encv{} spaces and has already been noticed in different works, see \cite{Gayral2007fk} for an example. It is often mentioned that the studied models are renormalisable at one-loop order provided one adds such a term. Our work is the first study to all orders of  such a quantum field theory.

One the other side, a (\encv) model on a degenerate space could open a way towards \encv{} \emph{Minkowski} space. There already exists lots of works about quantum field theory on \encv{} Minkowski space concerning mainly causality, unitarity, definition of the appropriate Feynman rules etc. But no renormalisable model is known. On commutative spaces, using a regularization \`a la Feynman, one can prove the perturbative renormalisability of a Minkowskian model from the corresponding Euclidean version. This results in the perturbative definition of the \emph{time-ordered} Green functions. Could we do an equivalent on a \encv{} space? It turns out that with a proper (i.e. preserving unitarity) definition of a time-ordered product on \encv{} Minkowski space \cite{Denk2003aa,Bahns2002aa}, one is lead to the conclusion that such a product is not equivalent to the use of a Feynman propagator. Therefore the usual techniques employed on commutative spaces may not apply. However one could address a simpler question namely the perturbative renormalisability of a field theory on a \encv{} Minkowski space with a commuting time\footnote{In that case, the modified Feynman rules reduce to the usual ones.}. This is where our present proposition could enter into the game.

\bibliographystyle{fabutphys-en}
\bibliography{biblio-articles,biblio-books}

\end{document}